\documentclass[floats,preprint,prl,aps]{revtex4}

\usepackage{graphicx}

\newcommand\beq{\begin{equation}}
\newcommand\eeq{\end{equation}}
\newcommand\bea{\begin{eqnarray}}
\newcommand\eea{\end{eqnarray}}
\newcommand{\nonum}{\nonumber}

\begin{document}

\title{\bf Magnetic flux induce dissipation effect on the  
quantum phase diagram of mesoscopic SQUID array  
}

\author{\bf Sujit Sarkar}
\address{\it
PoornaPrajna Institute of Scientific Research,
4 Sadashivanagar, Bangalore 5600 80, India.\\
}
\date{\today}
\begin{abstract}
We present the quantum phase diagram of mesoscopic SQUID array. 
We predict different quantum phases and the phase
boundaries along with several special points. We present the
results of magnetic flux induced dissipation on these 
points and also on the phase boundaries.
The Josephson couplings at these points are dependant on the system
parameter in a more complicated fashion 
and differ from
the Ambegaokar and Baratoff relation.
We derive the 
analytical relation between the dissipation strength and Luttinger liquid
parameter of the system. We observe some interesting behaviour at the
half-integer magnetic flux quantum and also the importance of
co-tunneling effect. \\


\end{abstract}

\maketitle


\section{ 1. Introduction}
\noindent
Josephson junction arrays have attracted considerable interest in the recent
years owing to their interesting physical properties 
like quantum phase transition, quantum critical behaviour and Coulomb blockade
etc \cite{schon,fazio}. The most universally observed behaviour for the low-dimensional 
superconducting system is the superconductor-insulator transition in superconducting
flim, wires, Josephson junction array in one and two dimensional giving rise
to intense experimental and theoretical activity
\cite{havi1,mason,bezr,bezr2,sondhi,lar,suj1,suj2,geer,zant,
havi3,kuo,gran,chen,refa}
This superconductor-insulator transition
occurs at low temperature ( mili-Kelvin scale) due the variation of one of the parameter 
of the system such as the normal state resistance of the flim, thickness of
the wire, the Josephson coupling of Josephson junction array and the magnetic 
flux of the mesoscopic SQUID array.\\
In one dimensional superconducting
quantum dot lattice system (mesoscopic SQUID array, Cooperpair box array) 
one elementary excitation occurs, i.e., the
quantum phase slip centers (QPS) \cite{suj2,kuo,chen,refa}. 
Recently the physics of QPS and the related
physics has been studied extensively. Quantum phase slip process is a
topological excitation, it's a discrete process in space-time domain of a
one dimensional superconducting system. In this process the amplitude of the
order parameter destroys temporarily at a particular point that leads the 
phase of the order parameters to change abruptly (in units of 2 $\pi$).
This process occurs at the time of macroscopic quantum tunneling of the
system. As we understand from our previous study \cite{suj2} and also from the
existing literature that the QPS process initiate the superconductor-
insulator transition \cite{chen,refa}. It is well known 
after the work of Calderia and Leggett \cite{cal}  
the dissipation plays a
central role in the macroscopic quantum system. 
There are several study in the literature based on the Calderia-Leggett
formalism to explain the different physical properties of low-dimensional
tunneling junction system \cite{sch1,su1,su2,schon2,kane,furu,zai1}.
We are mainly motivated from the experimental findings of 
Chow $et~al$ \cite{havi3}. They have done 
some pioneering work to study the effect of magnetic flux on
mesoscopic SQUIDs array. They have observed the magnetic flux
induced dissipation driven 
superconductor-insulator quantum phase transition and the
evidence of qunatum critical point. The other interesting part
of this study the length scale dependent superconductor-insulator
transition and the magnertic flux induced superconductivity.\\
The goal of this paper is to provide a detailed analysis of the
appearance of different quantum phases and phase boundaries in
the presence and absence of dissipation for a mesoscopic SQUID array at
$T= 0$. We assume the presence of
ohomic dissipation motivated by the experiment in Ref. (\cite{havi3} ) and
we study the consequence of it in the different quantum phases
and phase boundaries and also on the special points like the
particle-hole symmetric point, charge degeneracy point, and multicritical
points. We also
derive the relation between the dissipation strength and Luttinger
liquid parameter of the system. We are able to find out the
relation between the Josephson coupling and the other interaction parameters
of the system through the analysis of Luttinger liquid parameter of the system
, we will notice that this analytical relation is more complicated and
differs from the Ambegaokar-Baratoff \cite{ambe} relation. 
To the best of our knowledge,
this is the first derivation in the literature.
The plan of the manuscript is as follows.  In section 2A,
we introduce the basic concept of the appearence of quantum 
phase slip centers and the source of dissipation for one
dimensional superconducting quantum dot lattice. In section
2B, we derive the relation between the dissipation strength and
the Luttinger liquid parameter of the system. In section 3, We
present the analysis of quantum phases and phase boundaries, in
the presence and absence of magnetic flux induced dissipation. Section IV,
is devoted for summary and conclusions. 
\begin{figure}
\includegraphics[scale=0.35,angle=0]{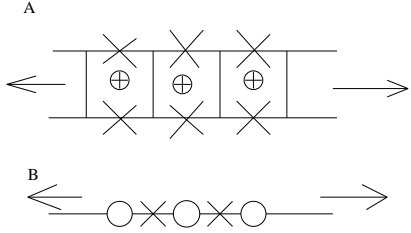}
\caption{A. Schematic diagram of one dimensional array of
small capacitance dc SQUIDs. Each plaque is a SQUID with
two Josephson junctions marked by the cross.
Circle with plus sign represent the applied magnetic flux $\Phi$.
B. Equivalent representation of system A, where the dots are
connected through tunnel junctions and the Josephson couplings
of this system is tunable due to presence of magnetic flux $\Phi$.}
\label{Fig. 1 }
\end{figure}
\vskip .2 true cm
\section{2. Basic Aspects of Quantum Phase Slip Centers and 
Analytical Derivations of Dissipative 
Strengths in Terms of Luttinger liquid parameter }
\noindent

\subsection{2A. Basic Aspects of Quantum Phase Slip Centers: }
It is well known to us that the Josephson coupling of the SQUID
is modulated by a facror, 
$E_{J1} = E_J |cos(\frac{\pi {\phi}}{{\phi}_0})|$, where $\phi$ is the
magnetic flux and ${\phi}_0 = \frac{hc}{2e}$ is the magnetic flux
quantum. Therefore we can consider the the mesoscopic SQUIDs array
as a superconducting quantum dot (SQD) lattice with modulated
Josephson junction \cite{suj1,suj2}. Fig. 1 shows the equivalance between the 
mesoscopic SQUIDs array and SQD lattice with modulated Josephson
junctions.
Here we prove the appearance
of QPS in SQD array with modulated Josephson coupling
through an analysis of a minimal model.
We consider two SQDs are separated by a Josephson junction.
These two SQDs are any arbitrary SQDs of the array. Appearance of
QPS is an intrinsic phenomenon (at any junction at any instant)
of the system. The
Hamiltonian of the system is
\beq
H~=~ \sum_{i} ~\frac{{n_i}^2}{2 C}~-~
{E_J} | cos ( \frac{\pi \Phi}{{\Phi}_0} ) | \sum_{i}
cos(~{\theta}_{i+1}~-~{\theta}_{i})
\eeq
where $n_i$ and ${\theta}_i $ are respectively the Cooper pair density and
the superconducting phase of the i'th dot and $C$ is the capacitance
of the junction.
The first term of the Hamiltonian present the Coulomb charging energies
between the dots and the second term is nothing but the Josephson
phase only term with modulated coupling, due to the presence of
magnetic flux. We will see that this model is sufficient to capture
the appearance of QPS in SQD systems. Hence this Hamiltonian has
sufficient merit to capture the low temperature dissipation physics
of SQD array.
In the continuum limit,
the partition function of the system is given by,
$
Z= \int  D \theta (x,\tau)  ~e^{- ~ {S_Q}  (x,\tau) },
$
where
$
S_{Q} = \int d \tau
\int dx \frac{E_J}{2} ~[{({\partial}_{\tau} \theta (x,\tau))}^2
 + {( {\partial}_x \theta(x,\tau) )}^2 ]
$.
This action is quadratic in scalar-field $\theta (x,\tau )$,
where $\theta (x,\tau )$
is a steady and differentiable field, so
one may think that no phase transition can occur
for this case. This situation changes drastically
in the presence of topological excitations for which $\theta (x,\tau)$ is
singular at the center of the topological excitations. So for this
type of system, we express the $\theta (x, \tau)$ into two components:
$\theta (x, \tau) = {{\theta}_0} (x, \tau) + {{\theta}_1} (x, \tau)$, where
${{\theta}_0} (x, \tau)$
is the contribution from attractive interaction of the system and
${{\theta}_1} (x, \tau)$ is the singular part from topological excitations.
We consider at any arbitrary time $\tau$, a topological excitation with
center at $X (\tau)~=~ ( {x_0} (\tau), {{\tau}_0} (\tau))$
. The angle is measured
from the center of topological excitations between the spatial coordinate
and the x-axis
$
{{\theta}_1} (x, \tau) ~=~ tan^{-1} (~\frac{{\tau}_0 - \tau}{x_0 -x} )
$
. The derivative of the angle is
$
{\nabla}_x {{\theta}_1} (x- X(\tau) )~=~ \frac{1}{{|x - X(\tau)|}^2}
 [-({\tau}_0 - \tau ), (x - x_0 )]
$
which has a singularity
at the center of the topological excitation.
Finally we get an interesting result when
we integrate along an arbitrary curve encircling
the topological excitations.
$
\int_{C1} dx  {\nabla}_x  {{\theta}_1} (x - X(\tau) )~=~ 2 \pi .
$
So we conclude from our analysis that when a topological
excitation is present in the SQD array, the phase difference
,$\theta$, across the junction of quantum dots jumps by
an integer multiple of $2 \pi$. This topological excitation is
nothing but the QPS in the $(x,\tau)$ plane.
According to the phase voltage Josephson relation
, $V_J~=~ \frac{-1}{2e} \frac{d \theta}{dt}$ ($t$ is the time),
a voltage drop occurs during this phase slip, which is the source
of dissipation.\\ 
\subsection{2B. 
Analytical Derivations of Dissipative 
Strengths in Terms of Luttinger Liquid Parameters }

Here we derive analytical expression of magnetic flux induced 
dissipative strength (${\alpha} $) in
terms of the interactions of the system. 
At first we derive the dissipative action/partition function 
of a
quantum impurity system. 
In our study, we consider the backscattering process from a quantum
impurity to compare the effective action with the superconducting
tunnel junction system.
We will see that the analytical structure of this
dissipative action is identical with the mesoscopic SQUIDs array.
In a different context, the authors of Ref. \cite{caza} have predicted the dissipation
driven quantum phase transition to occur. They have also considered the presence
of backscattering events originated in the LL under the effect of dynamically 
screened Coulomb interaction. 
\\
Here we consider that the impurity is present at the origin where the fermions
scatter from the left to the right and vice versa. The Hamiltonian describing
this process is
$$ {H_1} =  {V_0} ( {R^{\dagger}} (0) {L} (0) + h.c ) 
= {V_0} \int dx {\delta} (x) cos{\theta} (x) . $$ 
The total Hamiltonian of the system is
\beq 
 H = {H_0}~+~ {V_0} \int dx  {\delta} (x) cos{\theta} (x,\tau)
\eeq
\beq
 {H_0} = \frac{1}{2 \pi} \int { u K {({{\partial}_x} {\theta}(x,\tau) )}^{2} 
~+~ \frac{u}{K}
{({{\partial}_x} {\phi}(x,\tau ) )}^{2} }, 
\eeq
The corresponding Lagrangian of the system is
\bea
{L} & = & \frac{1}{2 \pi K} \int { \frac{1}{ u} 
{({{\partial}_{\tau}} {\phi}(x,\tau) )}^{2} ~+~ {u}
{({\partial} {\phi}(x,\tau) )}^{2} } \nonum\\ 
& & + {V_0} \int dx  \delta (x) cos(\theta (x,\tau)) 
= {L_0} + {L_1} 
\eea
where ${L_0}$ and ${L_1}$ are the non-interacting and the interacting part
of the Lagrangian and $K$ is the Luttinger liquid parameter of the system.
The only non-linear term in this Lagrangian is expressed by the field 
$\theta (x=0 )$. We would like to express the action of the system as an
effective action by integrating the field $\theta (x \neq 0 )$. Therefore
one may consider $\theta (x \neq 0 )$ as a heat bath, which yields the
source of dissipation in the system. The constraint condition 
for the integration is
$ {\theta} ( \tau ) = {\theta} (x=0, \tau )$. 
We can write the partition function of the quantum impurity system.
$ Z= \int  D \theta (x,\tau)  D \theta (\tau)
\delta (\theta (\tau) - {\theta} (0, \tau ) ) e^{- \int_{0}^{\beta} L d {\tau}} $
Now we use the standard trick of introducing the Lagrange multiplier with
auxiliary field $\lambda (\tau)$.
\bea
 Z &= & \int {D \theta (\tau)} e^{- \int_{0}^{\beta}  {L_1}  d {\tau} }
\int D {\lambda (\tau)} e^{-i {\lambda (\tau)} {\theta} (\tau )} \nonum\\  
 & &  \int {D \theta (x, \tau)} e^{ \int_{0}^{\beta}  (- { L_0} 
+ i {\lambda} ({\tau}) {\theta} (0, \tau) ) {d \tau} } 
\eea
The Fourier transform of the first term of Eq.(3) is
$ {L_0} ~=~  \sum_{q} \sum_{i {\omega}_n } 
\frac{ {{\omega}_n}^{2}~+~ {v}^{2} {q}^{2} }{2 \pi K v} 
{\theta} (q, i {{\omega}_n} ) {\theta} (- q, - i {{\omega}_n} ) $. 
At first we would like to calculate the integral: 
$ \int_{0}^{\beta} d {\tau} [{L_0}
- i {\lambda (\tau)} {\theta (0, \tau)} ] $, 
we can write this term as
\bea
\sum_{q} \sum_{i {\omega}_n }
\frac{ {{\omega}_n}^{2}~+~ {v}^{2} {q}^{2} }{2 \pi K v}
{\theta} (q, i {{\omega}_n} ) {\theta} (- q, - i {{\omega}_n} ) \nonum\\
 -\frac{1}{2 \sqrt{L} } ( {\lambda} (i {\omega}_{n}) 
 {\theta} (- q, - i {{\omega}_n}) + {\lambda} (- i {\omega}_{n})
 {\theta} (q, i {{\omega}_n}).
\eea
This integral appears in the integral $ \theta (x, \tau)$. This integral
is quadratic in $\theta $. Now we would like to perform the Gaussian 
integration by completing the square. We can write the result as
$ \frac{-1}{2 L} \sum_{i {{\omega}_n}, q} \frac{\pi K v}
{ {{\omega}_n}^{2}~+~ {v}^{2} {q}^{2} } $.
In the infinite length limit one can write,\\
$ \frac{1}{2 L} \sum_{q} \frac{\pi K v}
{ {{\omega}_n}^{2}~+~ {v}^{2} {q}^{2} }~=~ 
\int \frac{dq}{2 \pi} \frac{\pi K v}
{ {{\omega}_n}^{2}~+~ {v}^{2} {q}^{2} } = \frac{\pi K}{4 {{\omega}_n} } $.\\
Now we would like to append this result of integration in the 
second integral
of $Z$, i.e., the integral over ${\lambda}$. One can write the integrand 
as
\bea 
\sum_{i {{\omega}_n } } ( - \frac{\pi K}{4 {{\omega}_n} } 
{\lambda} (i {\omega}_{n}) {\lambda} (-i {\omega}_{n}) 
 + \frac{i}{2} ( {\lambda} (i {\omega}_{n})
 {\theta} (- q, - i {{\omega}_n}) \nonum\\
+ {\lambda} (- i {\omega}_{n})
 {\theta} (q, i {{\omega}_n}) .
\eea 
This integral is again the quadratic integral of $\lambda $, therefore, the
Gaussian integral can be performed by completing the square. We perform 
the integration and finally obtain
$ \sum_{i {{\omega}_n } } \frac{{\omega}_n}{\pi K} 
\theta (i {\omega}_{n}) \theta (- i {\omega}_{n}) $.
From this analytical expression, 
we obtain the effect of bath on ${\theta} (\tau) $. Appearance
of the factor $ {{\omega}_n} $ signifies the dissipation. 
Therefore, the effective
action reduces to
\beq
S ~=~ \sum_{i {{\omega}_n } } \frac{{\omega}_n}{\pi K}
\theta (i {\omega}_{n}) \theta (- i {\omega}_{n}) +  
\int d{\tau} {V_0}  cos{\theta} (\tau)  
\eeq 
The above action implies that a single particle moves in the potential
$ {V_0} cos{\theta} (\tau)$ subject to dissipation with friction constant
, $\frac{1}{\pi K}$.  

Now we calculate the dissipative action of mesoscopic SQUIDs
array. 
Here, we calculate 
the effective partition function of our system. Our
starting point is the Calderia-Legget \cite{cal} formalism.
Following this reference we write the action as 
\beq
S_1 ~=~ S_0 ~+~ \frac{{\alpha}}{4 \pi T}~ \sum_{m} {{\omega}_m}
{|{\theta}_m|}^2 .
\eeq
$S_0 $ is the action for non-dissipative part and
$S_1$ is the standard action for the system with tilted wash-board
potential \cite{sch1,su1,schon}
to describe the dissipative physics for low dimensional superconducting
tunnel junctions, ${\alpha} $ is the dissipative strength of the
system,
\beq
{\alpha} ~=~ \frac{R_Q}{R_s} cos|\frac{\pi \phi}{{\phi}_0}| 
\eeq
(the
extra cosine factor which we consider in ${\alpha}$ is entirely new 
which probes the effect of an external magnetic flux and
is also consistent physically),
${\omega}_m ~=~ \frac{2 \pi}{\beta} m$ is the Matsubara frequency 
and $R_Q$
($ = 6.45 k \Omega$) is the quantum resistance, $R_s$ is
the tunnel junction resistance, $\beta$ is the inverse temperature.
In one of our previous work, we have explained few experimental findings 
by using the above expression of ${\alpha} $. 
In the strong potential, tunneling between the minima
of the potential is very small.
In the imaginary time path integral formalism,
tunneling effect in the strong coupling limit can be described
in terms of instanton physics.
In this
formalism, 
it is convenient
to characterize the profile of $\theta$ in terms of its time derivative,
\beq
\frac{d \theta {( \tau)} }{d {\tau} }~=~\sum_{i} e_i h (\tau - {\tau}_i),
\eeq
where  $h (\tau - {\tau}_i)$ is the time derivative at time $\tau$ of
one instanton configuration.
${\tau}_i$ is the location of the i-th instanton, $e_i = 1$ and $-1$
is the topological charge of instanton and anti-instanton respectively.
Integrating the function over $h$ from
$-\infty$ to $\infty$, we get
$ \int_{-\infty}^{\infty} d \tau h (\tau) = {\theta} (\infty)
- {\theta}({-\infty}) = 2 \pi. $
It is well known that the instanton (anti-instanton)
is almost universally constant except for a very small region of
time variation.
In the QPS process the amplitude of the superconducting
order parameter is zero only in a very small region of space as a function
of time and the phase changes by $\pm 2\pi$. 
So our system reduces to a neutral system consisting of equal number of
instanton and anti-instanton.
One can find the expression
for ${\theta} ( \omega)$ after the Fourier transform applied to both 
the sides
of Eq. 11 which yields
$ {\theta} ( \omega ) = \frac{i}{\omega} \sum_{i} e_i h (i \omega)
e^{i {\omega} {\tau}_1 } $
. Now we substitute this expression for ${\theta} ( \omega )$ in the
second term of Eq. 9 and finally we get this term as
$ \sum_{ij} ~F({\tau}_i  - {\tau}_j ) {e_i} {e_j} $, where
$ F ({\tau}_i  - {\tau}_j ) = \frac{\pi \alpha}{\beta }
\sum_{m }~ \frac{1}{{ |{\omega}_m}|} e^{i {\omega} ({\tau}_i  - {\tau}_j )} $
$ \simeq  ln ({\tau}_i - {\tau}_j )$.
We obtain this expression for very small values of
${\omega}$ ( $\rightarrow 0 $).
So $ F ({\tau}_i  - {\tau}_j )$ effectively represents the Coulomb
interaction between the instanton and anti-instanton.
This term is the main source of dissipation of SQUID array 
system. Following
the standard prescription of imaginary time path integral formalism, we can
write the partition function of the system as
\cite{suj2,lar,gia1,kane,zai1,furu}.
\bea
Z & = & \sum_{n=0}^{\infty} \frac{1}{n !} 
{z}^{n} \sum_{e_i} \int_{0}^{\beta} d {{\tau}_n}
\int_{0}^{{\tau}_{n-1}}d {\tau}_{n-1} ... \nonum\\
& & \int_{0}^{{\tau}_2} d {{\tau}_1}
e^{-F ({\tau}_i  - {\tau}_j ) {e_i}{e_j} }.
\eea
We would like to express the partition function in terms of integration over
auxiliary field, $ q{( \tau )}$. After some extensive analytical calculations, 
we get 
\beq
Z = \int D q ( \tau ) e^{(- \sum_{i {\omega}_n }
 \frac{ |{ {\omega}_n}|}{4 \pi {\alpha} } 
q (i {\omega}_{n}) q (- i {\omega}_{n}) )
 {+ (2 z \int_{0}^{\beta} d {\tau} cosq ( \tau ) )} } .
\eeq
Thus by comparing the first term of the action of Eq. (8) and the first term of 
exponential of Eq. (13), we conclude 
that the dissipative strength $\alpha$ and the Luttinger liquid parameter
of the system are related by the relation, $K = 4 {\alpha} $. 
\\
\begin{figure}
\includegraphics[scale=0.45,angle=0]{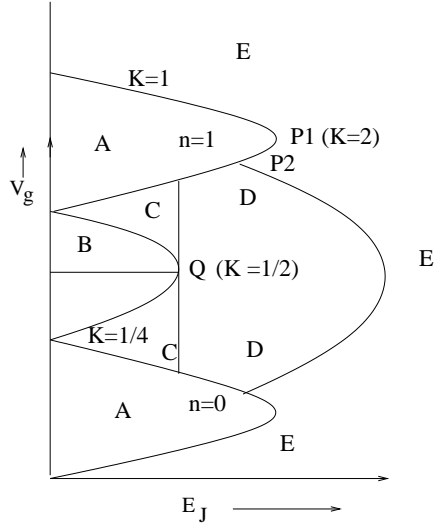}
\caption{
Quantum phase diagram ($E_{J}$ vs. $V_g$) of
the SQD array. We
have depicted the different phases of the model Hamiltonians by:
A. Mott insulating phase, B. Charge-density wave (CDW); 
C. First kind of Repulsive
Luttinger liquid (RL1); D. Second kind of Repulsive Luttinger liquid (RL2); 
E. Superconductivity. Q and
P2 are the multi-critical points (please see the text). $K$ is
1 at the phase boundary between Luttinger liquid (RL2) and superconductivity
and also at the phase boundary between Mott phase and superconductivity.
$K =1/4$ at the phase boundary between RL1 and CDW state. $K=2$ and $1/2$
at the $P1$ and $Q$ point respectively. $Q$ and $P1$ are the special points,
$Q$ is the charge degeneracy point and $P1$ is the particle-hole
symmetric point.
}
\end{figure}
\begin{figure}
\includegraphics[scale=0.45,angle=0]{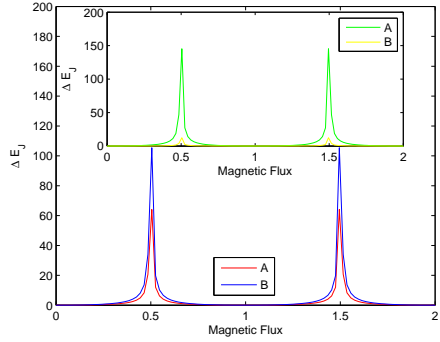}
\caption{Shift of the Josephson coupling ($\Delta E_J $) due
to the magnetic flux induced dissipation effect with magnetic flux
(measured with respect to magnetic flux quantum, ${{\phi}_0} = \frac{hc}{2e}$).
The red line is for the shifting of the Josephson coupling for the phase 
boundary between the 
RL2 and the superconducting phase. Blue line is for the
shift of the phase boundary between the CDW and RL1 phase. Inset shows the
shift of the Josephson coupling for the particle-hole symmetric point (green line)
and the Charge degeneracy point (yellow line) in the figure. 
}
\end{figure}
\begin{figure}
\includegraphics[scale=0.45,angle=0]{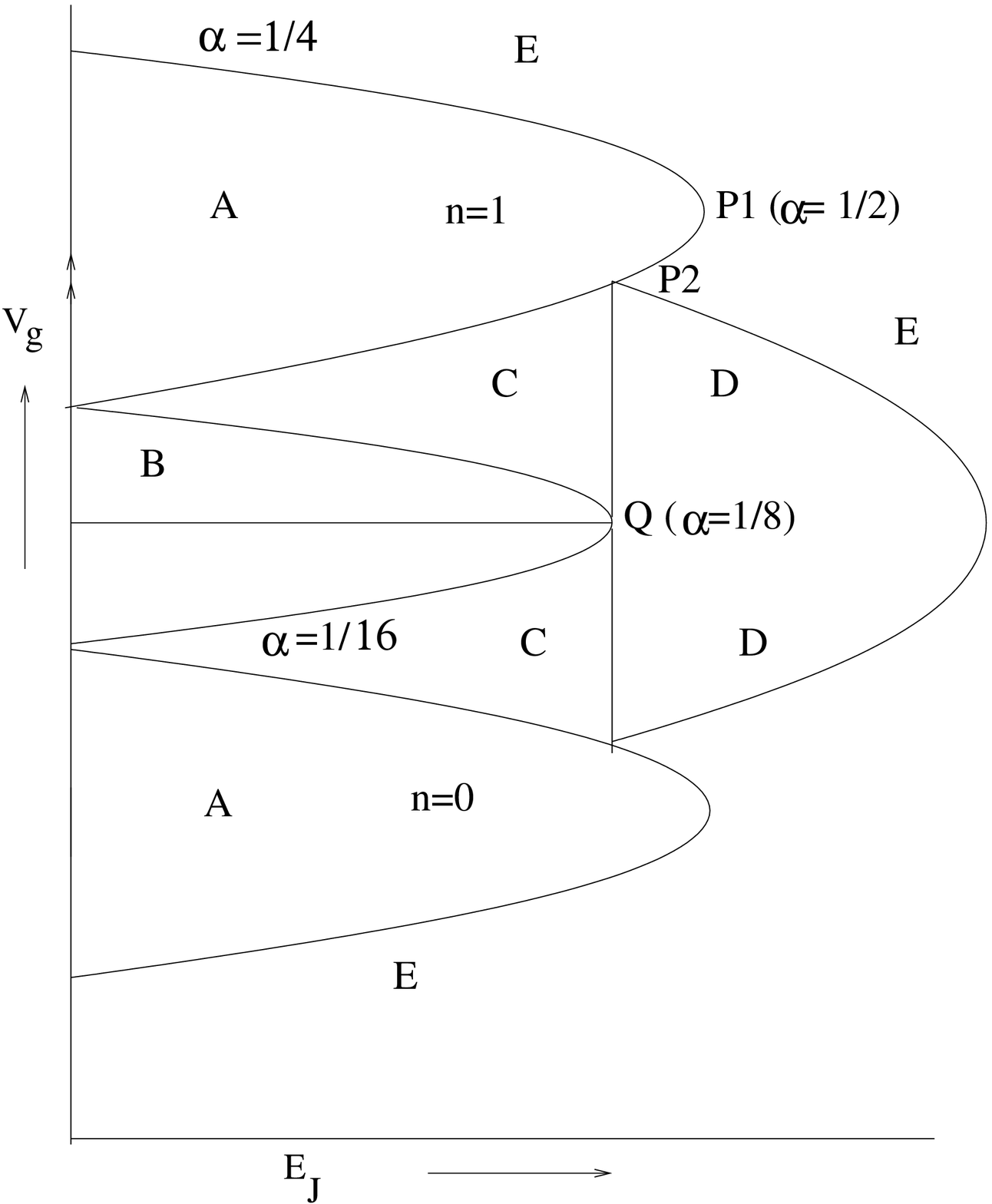}
\caption{Dissipative
quantum phase diagram ($E_{J}$ vs. $V_g$) of
the SQD array. We
have depicted the different phases of the model Hamiltonians by:
A. Mott insulating phase, B. Charge-density wave (CDW); 
C. First kind of Repulsive
Luttinger liquid (RL1); D. Second kind of Repulsive Luttinger liquid (RL2); 
E. Superconductivity. Q and
P2 are the multi-critical points (please see the text). $K$ is
1 at the phase boundary between Luttinger liquid (RL2) and superconductivity
and also at the phase boundary between Mott phase and superconductivity.
$K =1/4$ at the phase boundary between RL1 and CDW state. $K=2$ and $1/2$
at the $P1$ and $Q$ point respectively. $Q$ and $P1$ are the special points,
$Q$ is the charge degeneracy point and $P1$ is the particle-hole
symmetric point. We notice that due to the dissipation, phase boundaries and
special points are shifted w.r.t Josephson coupling, please see the text for
detail explanation.
}
\end{figure}
\section{ 3. Quantum Field Theoretical Study of Model Hamiltonian of The System
 and Explicit Derivation of Dissipative Strength}
In the previous section and also from our previous study \cite{suj2}, 
we have shown explicitly that the mesoscopic SQUIDs array
is equivalent to the array of superconducting quantum dots (SQD)
array with modulated
Josephson coupling.  
We first write the model Hamiltonian of SQD with nearest neighbor (NN)
Josephson coupling Hamiltonian ($H_{J1}$) and
also with the presence of the
on-site ($H_{EC0}$) and NN charging energy ($ H_{EC1}$) between SQD as,
\beq
H~=~H_{J1}~+~H_{EC0}~+~H_{EC1}.
\eeq
We would like to recast our model Hamiltonian in the magnetic
flux induced Coulomb blocked regime
(${E_{C0}} >> {E_{J}}$). In this regime, one can recast the
Hamiltonian in spin operators. 
It is also observed from the
experiments that the quantum critical point exists for 
larger values of the magnetic field, when the magnetic field
induced Coulomb blockade phase is more prominent than the
$E_J $ induced SC phase.
Thus our theoretical model
is consistent with the experimental findings.
During this mapping process we follow Ref. (\cite{lar,suj1,suj2}), so that.
$$
H_{J1}~=~ -2~E_{J1} \sum_{i}
( {S_i}^{\dagger} {S_{i+1}}^{-} + h.c)
$$, 
$ {E_{J1} }= {E_ {J} } |cos (\frac{\pi \Phi}{{\Phi}_{0}})| $,
$$
H_{EC0}~=~ \frac{E_{C0}}{2} \sum_{i}
{( {S_i}^{Z} -h )}^2.$$
are the Hamiltonian $H_{EC0}$ accounts for the
influence of gate voltage ($e N \sim V_g$), where
$e N$ is the average dot charge induced by the gate voltage.
When the
ratio $\frac{E_{J1}}{E_{C0}} \rightarrow 0$, the SQD
array is in the insulating state having a gap of the width
$\sim {E_{C0}}$, since it costs an energy $\sim E_{C0}$
to change the number of pairs at any dot. The exceptions are the
discrete points at $N~=~(2n+1)$, where a dot with charge $2ne$
and $2 (n+1) e$ has the same energy because the gate charge
compensates the charges of extra Cooper pair in the dot.
On this degeneracy point, a small amount of Josephson coupling
leads the system to the superconducting state.
Here $h = \frac{N - 2n - 1}{2} $ allows the tuning of the system around the
degeneracy point by means of gate voltage. 
$
H_{EC1}~=~4 E_{Z1} \sum_{i} {S_i}^{Z}~{S_{i+1}}^{Z}
$
. At the Coulomb blocked regime, the higher order expansion leads
to the virtual state with energies exceeding $E_{C0}$. 
In this second order process, the effective Hamiltonian reduces to
the subspace of charges $0$ and $2$, and takes the form
\cite{lar,suj1,suj2},
\beq
H_C ~=~- \frac{3 {E_{J1}}^2 }{4 E_{C0}} \sum_{i}
{{S_i}^Z}{{S_{i+1}}^Z} ~-~ \frac{{E_{J1}}^2}{E_{C0}}
\sum_{i} ({S_{i+2}}^{\dagger} {S_i}^{-} + h.c).
\eeq
With this corrections $H_{EC1}$
become
$$
H_{EC1}~\simeq~ (4 E_{Z1}~-~\frac{3 {E_{J1}}^2}{4 E_{C0}})
~\sum_i {{S_i}^Z}{{S_{i+1}}^Z}.
$$
In this analytical expression, we only consider the nearest 
-neighbor contribution of the interaction. There is no
evidence of next-nearest-neighbour interaction 
for mesoscopic SQUID array system \cite{havi3}.
One can express
spin chain systems to as spinless fermions systems through
the application of Jordan-Wigner transformation. 
In Jordan-Wigner transformation
the relation between the spin and the electron creation and
annihilation operators are
$ S^{z} (x)  =  \psi^{\dagger} (x) \psi (x) - 1/2 $, 
$ S^{-} (x) =  { \psi (x)} ~\exp [i \pi \sum_{j=-\infty}^{x-1} n_j]$,
where $S^{+} = { (S^{-})}^{+} ,  n (x) = {\psi^{\dagger}} (x) {\psi (x)} $
is the fermion number at
the site $x$.
We have
transformed all Hamiltonians in spinless fermions as follows:
$
H_{J1}~=~ -2~E_{J1} \sum_{i}
( {{\psi}_i}^{\dagger} {{\psi}_{i+1}}^{-} + h.c) $,
$
H_{EC0}~=~ 2 h {E_{C0}} \sum_{i}
({{\psi}_i}^{\dagger} {{\psi}_i} - 1/2).$
$ H_{EC1}~\simeq~ (4 E_{Z1}~-~\frac{3 {E_{J1}}^2}{4 E_{C0}})
~\sum_i ({{\psi}_i}^{\dagger} {{\psi}_i} - 1/2)
({{\psi}_{i+1}}^{\dagger} {{\psi}_{i+1}} - 1/2) .$
 
In order to study the continuum field theory of these Hamiltonians,
we recast the spinless
fermions operators in terms of field operators by a relation \cite{gia2}.
\beq
 {\psi}(x)~=~~[e^{i k_F x} ~ {\psi}_{R}(x)~+~e^{-i k_F x} ~ {\psi}_{L}(x)]
\eeq
where ${\psi}_{R} (x)$ and ${\psi}_{L}(x) $ describe the second-quantized
fields of right- and
the left-moving fermions respectively.
We would like to express the fermionic fields in terms of bosonic
field by the relation
\beq
 {{\psi}_{r}} (x)~=~~
\frac{U_r}{\sqrt{2 \pi \alpha}}~~e^{-i ~(r \phi (x)~-~ \theta (x))},
\eeq
where $r$ denotes the chirality of the fermionic fields,
right (1) or left movers (-1).
The operators $U_r$ preserve the anti-commutivity of fermionic fields. 
$\phi$ field corresponds to the
quantum fluctuations (bosonic) of spin and $\theta$ is the dual field of $\phi$. 
They are
related by the relations
$ {\phi}_{R}~=~~ \theta ~-~ \phi$ and  $ {\phi}_{L}~=~~ \theta ~+~ \phi$.
The Hamiltonians without ($ H_1 $) and with co-tunneling ($ H_2 $) effect
are the following 
\bea
H_1  & = & {H_0}
+ \frac{4 E_{Z1}}{{(2 \pi \alpha)}^2} \int ~dx :cos(4 \sqrt{K} \phi (x)):
\nonumber\\
& & + \frac{ E_{C0}}{\pi \alpha} \int  ({{\partial}_x} \phi (x))~dx
\eea
\bea
H_2  & = & {H_0}
+ \frac{( 4 E_{Z1} - \frac{3 {E_{J1}}^2}{4 E_{C0}})}{{(2 \pi \alpha)}^2}  
\int ~dx :cos(4 \sqrt{K} \phi (x)):
\nonumber\\
& & + \frac{ E_{C0}}{\pi \alpha} \int  ({{\partial}_x} \phi (x))~dx
\eea

Where, $H_0$ 
is the non-interacting part of the Hamiltonian. The Luttinger liquid parameters
of the Hamiltonian $H_1 $ and $H_2$ are $K_1 $ and $K_2 $ are respectively.
\beq
 {K_1} ~  =~  
\frac{\pi }
{\pi ~+~2 sin^{-1} {\Delta}_1 } 
\eeq
\beq
 {K}_2  ~ =~  \frac{ \pi }
{\pi~+~2 sin^{-1} {\Delta}_2} 
\eeq
${{\Delta}_1}= \frac{2 E_{Z1}}{E_{J1}} $
${{\Delta}_2}= \frac{2 E_{Z1}}{E_{J1}}~ -~\frac{3 {E_{J1}} }{8 E_{C0}} $.
We calculate the dissipation strength by calculating K for both cases and
then we use the relation $K = 4 {\alpha}$. Therefore the dissipative
strength in absence (${{\alpha}_1} $) and presence (${{\alpha}_2} $) 
of co-tunneling effect
are 
\beq
 {{\alpha}_1} ~  =~ \frac{1}{4} 
\frac{\pi}{\pi ~+~2 sin^{-1} {\Delta}_1 } 
\eeq
\beq
 {{\alpha}_2}  ~ =~\frac{1}{4} 
\frac{\pi}{\pi ~+~2 sin^{-1} {\Delta}_2 } 
\eeq
This is the first analytical derivation of flux induced dissipation strength
in terms of the interactions of the system. 
In the derivation of $K_2$ and ${\alpha}_2 $, we only consider the nearest-
neighbour hopping consideration when we consider the effect co-tunneling. 
We consider these two processes to
emphasis the importance of co-tunneling effect for this system.
Before we proceed further for the analysis of the Hamiltonian
, $H_1$ and $H_2$, we want to explain in detail for the different
values of $K$ at the different points (like P1, Q and P2) of phase diagram and also
at the phase boundaries. The physical analysis of the phases and
clear distinction of the phase boundaries will depend on the values of the
K.
Point Q is the charge degeneracy point for the
low Cooper-pair density ($n_i =1/2$). The value of $K$ will be evaluated
from the relevance of sine-Gordon terms. At the point Q, system is in
the second order commensurability, sine-Gordon coupling terms
,Eq. 18 and Eq. 19, will
become relevant for $K= 1/2$, which is depicted in the Fig.1.
Point P1 is the particle hole symmetric point,
i.e., there is one particle in each site. At this point system is in the
first order commensurability, i.e., the sine-Gordon coupling term
is $cos (2  \sqrt{K} \phi (x))$. So this term will become relevant
for K=2, which is depicted in the Fig.1.
We are understanding from Eq. 18 and Eq. 19 that
the applied gate voltage acts as a chemical potential. So the
proper tuning of gate voltage will drive the system from the
insulating state to the other quantum phases of the system. We are now interested
in finding the value of K at the phase boundaries, hence analysis is the
following: We follow the Luther-Emery \cite{luth} trick during the analysis.
One can write the sine-Gordon Hamiltonian for arbitrary
commensurability as
\beq
H_3 ~=~ H_0 ~+~ \lambda \int~ dx~cos (2 n \sqrt{K} \phi (x) ),
\eeq
where $n$ is the
commensurability and $\lambda$ is the coupling strength.
$H_0$ is the free part of the Hamiltonian.
We know that for the spinless fermions,
${{\psi}_R}^{\dagger}{{\psi}_L}~+~{{\psi}_L}^{\dagger}{{\psi}_R}
~=~\frac{1}{{2 \pi a}^2} \int dx cos (2 \sqrt{K} \phi (x)) $, which is
similar to the analytical expression of sine-Gordon coupling term
but with the wrong coefficient inside the cosine. One can set
${\tilde {\phi} (x)}~=~ 2 \sqrt{\tilde{K}} \phi (x)$ then the Eq. 24 become
\beq
H_4 ~=~ H_0 + \lambda \int~dx cos(2 {\tilde {\phi} (x)}).
\eeq
$K$ and $\tilde{K}$ are related by the relation, $ K~=~\frac{\tilde{K}}{n^2}$.

At the phase boundary, $\tilde{K}
~=1$ that implies $K=1/n^2$. So for the first and second
order commensurability the value of $K$ at the phase boundary are
1 and 1/4 respectively which is depicted in Fig. 1.
The point to be noticed that if we start from
an initial model with $K~=~1/n^2$, i.e., in general a strongly 
interacting model, the resulting spin-less fermions model corresponds 
in the boson language
to $K=1$ which means that it is non-interacting. For this particular
value of $K$ the spin-less fermions whose bosonized form is Eq. 25
are just free particle with backscattering. This special value of $K$ is
known as the Luther-Emery \cite{luth} line, the importance of the
Luther-Emery solution is to provide a solution for the massive
phase on the whole line $K= 1/n^2$ for arbitrary $\lambda$.
\\
Here we do the  
analysis for Hamiltonian $H_1$. In the limit $\Delta~=~{\Delta}_1$,
for $E_{J} < 2 E_{Z1}$
and
relatively small field, the anti-ferromagnetic  Ising interaction
dominate the physics of anisotropic Heisenberg chain. When the 
field is large, i.e., the
applied gate voltage is large, 
the chain state is in the ferromagnetic state.
In the language of interacting bosons, The Neel phase is the 
commensurate charge density wave phase with period 2, i.e.,
there is only one boson in every two sites. 
In Fig. 1 this phase region is described by region B. 
The ferromagnetic state is 
the Mott insulating state, this is the
phase A of our quantum phase diagram (Fig. 1).
$H_1$ is the Heisenberg XXZ model Hamiltonian in a magnetic
field. The emergence of two Luttinger liquid phases for the
following reasons: RL1 and RL2 respectively occur due to
commensurate-incommensurate transition and the criticality of
Heisenberg XY model. 
For the intermediate values of the
field, system is either in the 
first kind of
repulsive Luttinger liquid (RL1) phase
for $K < 1/2$ or in the 
second kind of repulsive Luttinger liquid (RL2) phase for $K > 1/2$.
The physical significance of RL1 phase
is that the coupling term is relevant but the applied magnetic
field, i.e., the applied gate voltage on the dot,
breaks the gapped phase whereas in the RL2 
phase non of the coupling term is relevant due to
the larger values of $K$ ($>1/2$). 
The phase regions are described by C and D 
respectively for the RL1 and RL2 in
Fig.2.  
These phase regions in Fig. 2,
are all most same for two different limits, 
$\Delta = {\Delta}_1$ and $\Delta = {\Delta}_2$.
So we predict the existence of two RL 
from two different sources. In previous studies this clarification
was absent and they had reported only one RL \cite{lar}.
The value of $K =1$ at the phase boundary between the MI and SC
phase and also between the RL2 and SC phase. From the analysis
of $K$ at the phase boundary we obtain $E_{Z1} =0 $, according 
to our theory and also from the experimental findings 
this condition is unphysical.
So the interaction space of Hamiltonian 
$H_1$ is not sufficient to produce the whole phase
diagram of Fig. 2. It indicates that we shall have to consider
more extended interaction space to get the correct phase diagram.
If we consider the co-tunneling effect in this Hamiltonian system,
i.e., the Hamiltonian $H_2 $. The phase boundary analysis at the
MI and SC phase and also for the RL2 and SC phase implies that
we get the condition 
$E_{J1}~=~ \sqrt{\frac{16 E_{Z1} E_{C0}}{3} }$, which is consistent
physically.  
Now we discuss the effect of magnetic field induced dissipation on the
quantum phase diagram of superconducting quantum dot lattice. We have
already proven in the previous section 
the analytical relation between the LL parameter ($ K$) and
the dissipation strength in presence of magnetic flux. Here we use the analytical
expression for $K$ to study the effect of magnetic flux on the quantum
phases and quantum phase boundaries. Therefore the modified quantum
phase diagram in presence of magnetic flux include the 
magnetic flux induce dissipation effect. 
In the previous paragraph we emphasis the importance of co-tunneling
effect. Therefore we consider the Eq.(21) and Eq.(23) when we consider
the shift of the Josephson couplings due to the presence of magnetic
flux induced dissipation. It is very clear from Eq. 21 and Eq. 23 that
the analytical relation of the Josephson coupling with the system interactions
parameters at different phase boundaries and different points are more
complicated than the Ambegaokar and Baratoff relation \cite{ambe}.
The analytical expression for the Josephson couplings at the different
phase boundaries and special points are the following:\\
Particle-Hole symmetric  point ($K =2$),\\
\beq
E_J ~ =~0.9428 {E_{C0}} 
( \sqrt{ 1 + \frac{6 E_{Z1}}{ E_{C0}}} +1) 
\eeq  
Charge-Degeneracy point $(K =1/2)$,\\
\beq
E_J ~ =~ {0.75 E_{C0}}
( \sqrt{ 1 + \frac{3 E_{Z1}}{E_{C0}}} -1) 
\eeq  
Phase boundary between the RL2 and SC phase $(K =1)$,\\
\beq
E_J ~ =~ {4}
( \sqrt{\frac{E_{Z1} E_{C0} }{3}} ) 
\eeq  
Phase boundary between CDW state RL1 phase $(K =1/4)$,\\
\beq
\Delta E_J ~ =~ {0.75 E_{C0}}
( \sqrt{ 1 + \frac{3 E_{Z1}}{ E_{C0}}} +1). 
\eeq 
Our main intension is to
study the effect of magnetic flux on the particle hole symmetric point ($K=2$ ),
charge degeneracy point ($K=1$), multicritical point and also for the phase boundaries between
the different quantum phases. \\
The analytical expressions for the shift of
Josephson couplings ($ \Delta E_J $)  
due to the presence of magnetic flux are the following:\\
Particle-Hole symmetric  point ($\alpha =1/2$),\\
\beq
\Delta E_J ~ =~0.9428 {E_{C0}} 
( \sqrt{ 1 + \frac{6 E_{Z1}}{ E_{C0}}} +1) 
(1/|cos(\frac{\pi \phi}{ {\phi}_0 }) | -1 )
\eeq  
Charge-Degeneracy point $(\alpha =1/8)$,\\
\beq
\Delta E_J ~ =~ {0.75 E_{C0}}
( \sqrt{ 1 + \frac{3 E_{Z1}}{E_{C0}}} -1) 
(1/|cos(\frac{\pi \phi}{ {\phi}_0 }) | -1 )
\eeq  
Phase boundary between the RL2 and SC phase $(\alpha =1/4)$,\\
\beq
\Delta E_J ~ =~ {4}
( \sqrt{\frac{E_{Z1} E_{C0} }{3}} ) 
(1/|cos(\frac{\pi \phi}{ {\phi}_0 }) | -1 )
\eeq  
Phase boundary between CDW state RL1 phase $(\alpha =1/16)$,\\
\beq
\Delta E_J ~ =~ {0.75 E_{C0}}
( \sqrt{ 1 + \frac{3 E_{Z1}}{ E_{C0}}} +1) 
(1/|cos(\frac{\pi \phi}{ {\phi}_0 }) | -1 )
\eeq 
In Fig. 3, we present our results of
deviation of Josephson coupling as a function of magnetic flux. We observe that this
deviation is maximum at the half-integers values of magnetic
flux quantum. It reveals from our study that the effect of 
magnetic flux induce dissipation for the half-integer values of
magnetic flux quantum is slightly prominant for  
the special points compare to the qunatum phase
boundaries. Inset shows the shift of Josephson
coupling for the particle-hole symmetric point and the charge
degeneracy point. It is also clear from the inset that the magnetic
flux induce dissipation is more prominant for the particle-hole
symmetric point compare to the charge degeneracy point.\\
It is very clear from the analytical expression from 
Eq. 30 to Eq. 33 and also from the Fig.3 from our study that
there is no appreciable changes 
in the quantum phase diagram for small magnetic flux in 
the system. As the applied magnetic fluxes changes from $0.4 {\phi}_0$
to $0.6 {\phi}_0$, quantum phase diagram shows some appreciable change 
and it is robust for the half-integer magnetic flux quantum.\\
In Fig. 4, we present the magnetic flux induce dissipative quantum phase diagram 
of our system. This schematic phase diagram is for values of magnetic flux
which appreciably effect the phase boundaries and special points as we have
discussed in the previous paragraph. We present the phase boundaries and
special points in terms of the dissipative strength of the system. We
observe from our study that magnetic flux induce dissipation favour
the insulating phase and the gapless LL phase over the superconducting 
phase of the system which is consistent with the experimental findings [13].\\   

\section{summary and conclusions}
We have studied the quantum phase diagram of mesoscopic SQUID array
in absence and presence of magnetic flux. The magnetic flux induced
dissipation modified quantum phase diagram 
of our system.
We have derived an analytical relation between the
Luttinger liquid parameter and dissipation strength. We have also 
noticed that magnetic flux induced dissipation effect is not same
for all values of magnetic flux quantum. We have also
observed that the magnetic flux induce dissipation favours the
insulating phase of the system over the Luttinger liquid and superconducting
phase of the system.\\    
{Acknowledgement: The author would like to acknowledge Dr. N. Sundaram
for reading the manuscript critically.}\\

\end{document}